\documentclass[aps,twocolumn,superscriptaddress,showpacs,longbibliography,pra,10pt]{revtex4-1}%
\usepackage{amsfonts}
\usepackage{amsmath}
\usepackage{amssymb}
\usepackage{graphicx}
\usepackage{dcolumn}
\usepackage{bm}
\usepackage{hyperref}
\usepackage{color}

\begin{document}

\title{Influence of the multiband sign changing superconductivity on the vortex cores and on vortex pinning in the new stoichiometric high {T$_c$} CaKFe$_4$As$_4$}
 
\author{Ant\'on~Fente}
	\affiliation{Laboratorio de Bajas Temperaturas y Altos Campos Magn\'eticos, Unidad Asociada UAM/CSIC, Departamento de F\'isica de la Materia Condensada, Instituto de Ciencia de Materiales Nicol\'as Cabrera and Condensed Matter Physics Center (IFIMAC), Universidad Aut\'onoma de Madrid, Spain}

\author{William~R.~Meier}
	\affiliation{Ames Laboratory, Ames, IA 50011}
	\affiliation{Department of Physics $\&$ Astronomy, Iowa State University, Ames, IA 50011}

\author{Tai~Kong}
	\affiliation{Ames Laboratory, Ames, IA 50011}
	\affiliation{Department of Physics $\&$ Astronomy, Iowa State University, Ames, IA 50011}

	\author{Vladimir~G.~Kogan}
		\affiliation{Ames Laboratory, Ames, IA 50011}

\author{Sergey~L.~Bud'ko}
	\affiliation{Ames Laboratory, Ames, IA 50011}
	\affiliation{Department of Physics $\&$ Astronomy, Iowa State University, Ames, IA 50011}

\author{Paul~C.~Canfield}
	\affiliation{Ames Laboratory, Ames, IA 50011}
	\affiliation{Department of Physics $\&$ Astronomy, Iowa State University, Ames, IA 50011}

\author{Isabel~Guillam\'on}
	\affiliation{Laboratorio de Bajas Temperaturas y Altos Campos Magn\'eticos, Unidad Asociada UAM/CSIC, Departamento de F\'isica de la Materia Condensada, Instituto de Ciencia de Materiales Nicol\'as Cabrera and Condensed Matter Physics Center (IFIMAC), Universidad Aut\'onoma de Madrid, Spain}

\author{Hermann~Suderow}

	\affiliation{Laboratorio de Bajas Temperaturas y Altos Campos Magn\'eticos, Unidad Asociada UAM/CSIC, Departamento de F\'isica de la Materia Condensada, Instituto de Ciencia de Materiales Nicol\'as Cabrera and Condensed Matter Physics Center (IFIMAC), Universidad Aut\'onoma de Madrid, Spain}
	
\date{\today}

\begin{abstract}
We study the superconducting density of states and vortex lattice of single crystals of CaKFe$_4$As$_4$ using a scanning tunneling microscope (STM). This material has a critical temperature of $T_c= 35\,$ K, which is one of the highest among stoichiometric iron based superconductors (FeBSC) and is comparable to $T_c$ found near optimal doping in other FeBSC. Using quasi-particle interference we identify the hole sheets around the zone center and find that two superconducting gaps open in these sheets. The scattering centers are small defects that can be localized in the surface topography and just produce quasiparticle interference, without suppressing the superconducting order parameter. This shows that sign inversion is not within hole bands, but between hole and the electron bands. Vortex core bound states show electron-hole asymmetric bound states due to proximity of the top of one of the hole bands to the Fermi level $E_F$. This places CaKFe$_4$As$_4$ in a similar situation as FeSe or related materials, with a superconducting gap $\Delta$ just a few times smaller than $E_F$. On the other hand, we also identify locations showing strong suppression of the superconducting order parameter. Their size is of order of the vortex core size and vortices are pinned at these locations, leading to a disordered vortex lattice.
\end{abstract}
 \maketitle

\section{Introduction.}

The Fermi surface of iron based superconductors (FeBSC) consists of many different sheets with different magnitudes of the superconducting gap. Often, gap magnitudes cluster around two values, leading to so-called effective two-gap superconductivity\cite{Kamihara08,Canfield10,Paglione10,Hirschfeld11,Efremov11,Hosono15}. The highest $T_c$'s of FeBSC are either found in doped systems with substitutional disorder or under substantial pressure, strain or stress, except in the recently discovered $AeA$Fe$_4$As$_4$ ($Ae=$Ca, Sr, Eu and $A=$K, Rb, Cs) systems \cite{Iyo_16, Paul1} where $T_c\cong 35\,$K is obtained in stoichiometric compounds. This provides an outstanding opportunity to understand the main features characteristic of superconductivity in FeBSC, because there is no substitutional disorder. In addition, there is no structural nor magnetic transition, which eliminates disorder due to domains of different crystalline orientations often found in FeBSC\cite{Paul1,PhysRevLett.118.107002}.

Penetration depth and tunneling experiments in CaKFe$_4$As$_4$ show evidence for two well defined superconducting gaps and sign changing behavior in agreement with s$_{\pm}$ superconductivity \cite{Prozorov16}. Results are comparable to those in optimally doped (Ba$_{1-x}$K$_x$)Fe$_2$As$_2$. The band-structure is also similar to that of (Ba$_{1-x}$K$_x$)Fe$_2$As$_2$, with three hole pockets at the $\Gamma$ point and two electron pockets at the $M$ point, having two gap values opening over the different Fermi surface sheets\cite{Prozorov16,Kaminski16}. NMR experiments show spin-singlet superconductivity and no Hebel-Slichter peak, with exponentially decaying spin relaxation rates, all compatible with nodeless s$_{\pm}$ superconductivity. Furthermore, alternating K and Ca layers are found to be well ordered and signatures of antiferromagnetic spin fluctuations are also found\cite{Cui17}. Substitution of Fe in CaKFe$_4$As$_4$ with Ni and Co, stabilizes a new kind of non-collinear magnetic order\cite{Meier17}. Here we take advantage of the unique properties of CaKFe$_4$As$_4$ to identify the Fermi surface sheets showing different superconducting gap values and order parameter sign changes and to determine the influence of the structure of the superconducting gap in vortex core and lattice.

\begin{figure*}[htb]
	\begin{center}
		\includegraphics[width=0.9\textwidth] {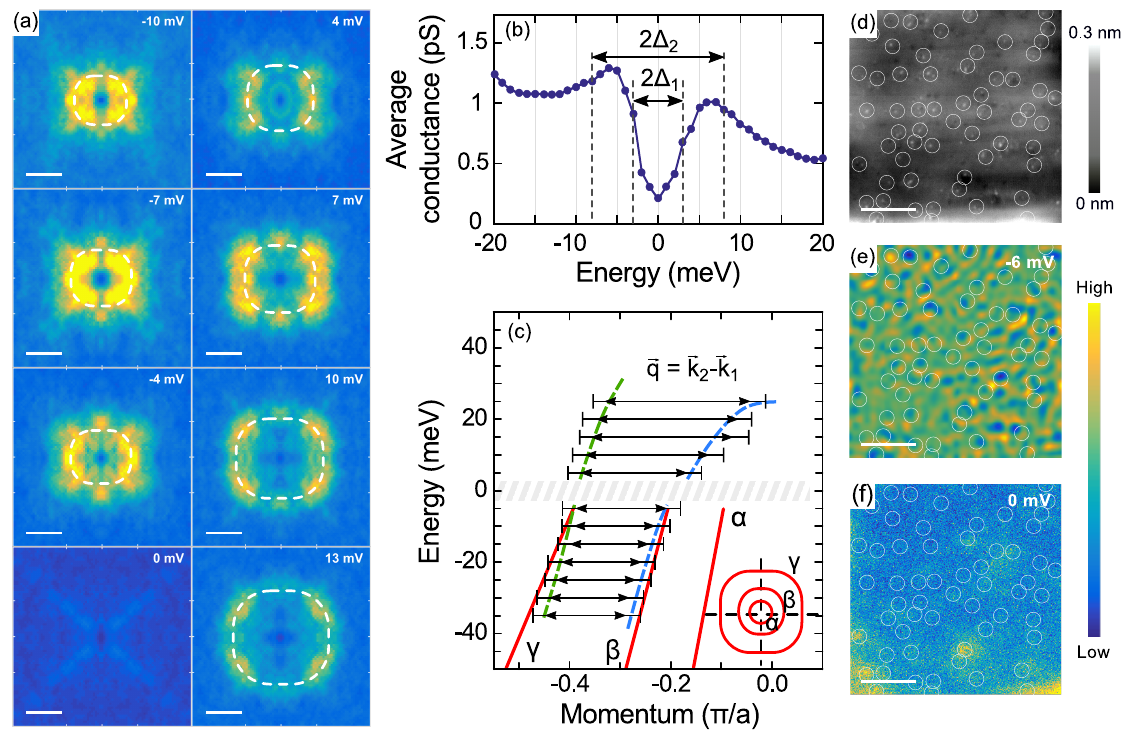}
		\vskip -.1 cm
		\caption{{\bf Quasiparticle interference and superconducting gap magnitude in CaKFe$_4$As$_4$.} (a) The Fourier transform of scattering patterns at defects at different bias voltages in zero magnetic field and 0.8 K. White scale bar is of 0.1 $\pi/a$ where $a$ is the in-plane lattice parameter ($a=3.861$\,\AA\,at about 6 K, see \protect\cite{Paul1}). (b) The magnitude of the Fourier transform averaged over the path marked by white dashed lines in (a) as a function of the bias voltage. (c) The reciprocal space vectors obtained from our data (green and blue dashed lines). Arrows provide the averaged radius of the interference signal shown in (a). Lateral error bars provide its width. We also plot the band dispersion obtained from ARPES as red lines, including $\alpha$, $\beta$ and $\gamma$ sheets. In the bottom right inset we show schematically the Fermi surface (hole pockets) of CaKFe$_4$As$_4$ obtained from ARPES. (d) The real space topographic STM map (color scale bar giving height differences at the right). White circles highlight defects observed at the surface. (e) and (f) Tunneling conductance maps roughly at the gap edge (e) and at zero bias (f) using the same conductance scale (color bar giving conductance differences at the right). White circles are at the positions where we find defects in (d). White scale bars in (d,e,f) are of 20 nm.}
		\label{f1}
	\end{center}
\end{figure*}

We study CaKFe$_4$As$_4$ using cryogenic scanning tunneling microscopy (STM)---a powerful tool offering insight into the superconducting and electronic material properties, due to an outstanding spatial and energy resolution. By applying a magnetic field, we can image the vortex lattice and connect electronic dispersion, gap structure and properties of the vortex core and the vortex lattice. We first determine the gap structure in different hole sheets of the Fermi surface using quasiparticle interference. We then examine the vortex core, relating the lowest Caroli-de Gennes-Matricon bound state with the shape of the hole bands close to the Fermi level. We also determine the vortex core size and its magnetic field dependence. Finally, we identify the vortex pinning mechanism, which is due to defects that suppress the order parameter at length scales of order of the vortex core size.

\section{Experimental.}

We grew single crystals of CaKFe$_4$As$_4$ from a high temperature Fe-As rich melt, using As ($99.9999\,\%$), K ($99.95\,\%$) and Fe ($99.9\,\%$) from Alfa Aesar, and Ca from Ames Laboratory Materials Preparation Center (99.9\%), following the procedure described in Ref.\cite{Paul1}. Crystals were screened to make sure that they are single phase, as in Refs.\cite{Paul1,Prozorov16,Kaminski16,Cui17,Meier17,Fente16b,PhysRevMaterials.1.013401} and have shiny flat surfaces. Crystals are plate-like, of several millimeters size and several hundreds of microns thick, with the $c$-axis perpendicular to the surface. We cleaved the samples \textit{in-situ} along (001) by gluing a brass stick to the surface and removing it at $4.2\,$K using a movable sample holder \cite{Suderow11,Fente16b}. After performing the experiment and warming the system, we saw that the cleaved surface is optically shiny and flat. The base temperature of our cryogenic system was of $800\,$mK and we used a gold tip prepared \textit{in-situ} as described in Ref.\cite{Rodrigo04}. By moving the sample holder below the tip, we can change the scanning window \textit{in-situ}. During this particular experiment, we have studied about 50 different fields of view 2$\times$2 $\mu$m in size. Magnetic fields are applied perpendicular to the surface (parallel to the $c$-axis) using a superconducting coil (in zero field cooled conditions). We report the tunneling conductance normalized by its value at a bias voltage far from the superconducting features.
CaKFe$_4$As$_4$ consists of atomically flat areas showing disordered arrangements of small size features (a few nm size). The surface looks similar to surfaces found in (Ba$_{0.6}$K$_{0.4})$Fe$_2$As$_2$\cite{Shan11}.

\section{Bandstructure and superconducting gap.}

Let us first discuss tunneling conductance maps at zero magnetic field. Real space tunneling conductance maps show a lot of structure at wavelengths larger than interatomic distances due to quasiparticle interference. By making maps with enough points in real space, we cover wavevectors in reciprocal space of order of the size of the Fermi surface pockets. We show Fourier transforms of real space conductance maps at different bias voltages in Fig.\,\ref{f1}(a). We mainly observe a roughly circular scattering feature whose radius in reciprocal space changes with bias voltage. Within the superconducting gap (lower left panel in Fig.\,\ref{f1}(a)), the scattering intensity strongly decreases.

The quasiparticle scattering intensity $g(E,\vec{q})$, with $\vec{q}$ the scattering vector from $\vec{k_1}$ to $\vec{k_2}$ ($\vec{q}=\vec{k_2}-\vec{k_1}$), is given by the scattering potential $V_S$ and the joint density of states $J(E,\vec{q})$, $g(E,\vec{q})\propto |V_S(\vec{q})|^2J(E,\vec{q})$. This assumes elastic scattering and neglects the energy dependence of the tunneling matrix elements and coherence effects due to superconductivity\cite{HoffmanPnictides,HIRSCHFELD2016197}. The joint density of states is given by  $J(E,\vec{q})\propto N_1(E,\vec{k_1})N_2(E,\vec{k_2})$ where $N_1$ and $N_2$ are the densities of states at $\vec{k_1}$ and $\vec{k_2}$. $J(E,\vec{q})$ is maximal for vectors $\vec{q}$ connecting two parts of the electronic dispersion on equal energy contours at the energy $E$ (measured with respect to the Fermi level, $E_F$). The tunneling conductance as a function of the bias voltage $V$ follows $g(\vec{q})$ as a function of $E$ (with $V=0$ for $E=E_F$). Thus, the tunneling conductance maps follow the electronic dispersion relation for filled ($V\le 0$) and empty ($V\ge 0$) states. The intensity of the scattering pattern is given by joint density of states $J(E,\vec{q})$ and the scattering potential $|V_S(\vec{q})|^2$.

The quasiparticle scattering pattern shown in Fig.\,\ref{f1}(a) is nearly circular although there are in-plane anisotropic features appearing at different bias voltages. These are due to asymmetry in the defects producing scattering, which we can ascribe to an anisotropic $V(\vec{q})$. The directionality is weak and there is a finite scattering intensity for all directions. Thus, we can safely take the joint density of states $J(E,\vec{q})$ corresponding to our quasiparticle pattern as being in-plane isotropic.

The radius of the scattering pattern in the Fourier transform gives a scattering wavevector $\vec{q}$ that coincides with the distance in reciprocal space between the hole bands around the $\Gamma$ point. In particular, between the $\gamma$ band and the $\beta$ bands. The considerable increase in $\vec{q}$ above the Fermi level shows that the $\beta$ band reaches its top very close to the Fermi level (Fig.\,\ref{f1}(c)).

In Fig.\,\ref{f1}(b) we show the Fourier transform of the scattering pattern averaged along the dashed line as a function of the bias voltage. This shows the joint density of states, $J(E)\propto N_1(E)N_2(E)$ at the vector corresponding to the scattering between both bands. The quasiparticle peak is located at a similar position as in previous measurements of the tunneling conductance vs bias voltage\,\cite{Prozorov16}. There are small shoulders at the values of the superconducting gap $\Delta_1$ and $\Delta_2$ determined using tunneling conductance and penetration depth\,\cite{Prozorov16}. The quasiparticle interference obtained here shows that $\Delta_1$ and $\Delta_2$ both open in the hole bands surrounding the zone center. This compares well with angular revolved photoemission (ARPES) measurements, which show the largest and smallest gap values at, respectively, the $\beta$ and $\gamma$ sheets\cite{Kaminski16}. Using quasiparticle intereference we can access the empty states and find that the $\beta$ band, which also carries the largest superconducting gap $\Delta_2$ according to ARPES, has its top very close to the Fermi level.

By comparing tunneling conductance with topographic images, we can discuss the relationship among different defects and the superconducting order parameter. In Fig.\,\ref{f1}(d) we show the topographic STM image obtained at the same time as the conductance maps. We have highlighted defect positions by white circles. In Fig.\,\ref{f1}(e) we show a representative tunneling conductance map within the gap, at $V=$\,-6\,mV. There are modulations with large wavelength whose intensity is mostly located close to the defects highlighted by the white circles. However, the zero bias conductance map (Fig.\,\ref{f1}(f)) remains unaffected by these scattering centers. This result implies that scattering between hole bands does not produce any suppression of the order parameter. This shows that the gap has the same sign in the bands close to zone center. The sign changing order parameter identified using penetration depth measurements in \cite{Prozorov16} must occur between the band at the zone boundary and the bands at the zone center.

In addition, we note that there are scattering centers in the field of view that lead to a suppression of the order parameter, around locations showing an increased zero bias conductance, see bottom part of Fig.\,\ref{f1}(f). These areas are quite large in size, reaching several nm. The scattering producing suppression of the order parameter does not contribute significantly to quasiparticle scattering patterns, but has a strong effect on vortex pinning, as we discuss below.

\begin{figure*}[htb]
	\begin{center}
		\includegraphics[width=0.9\textwidth] {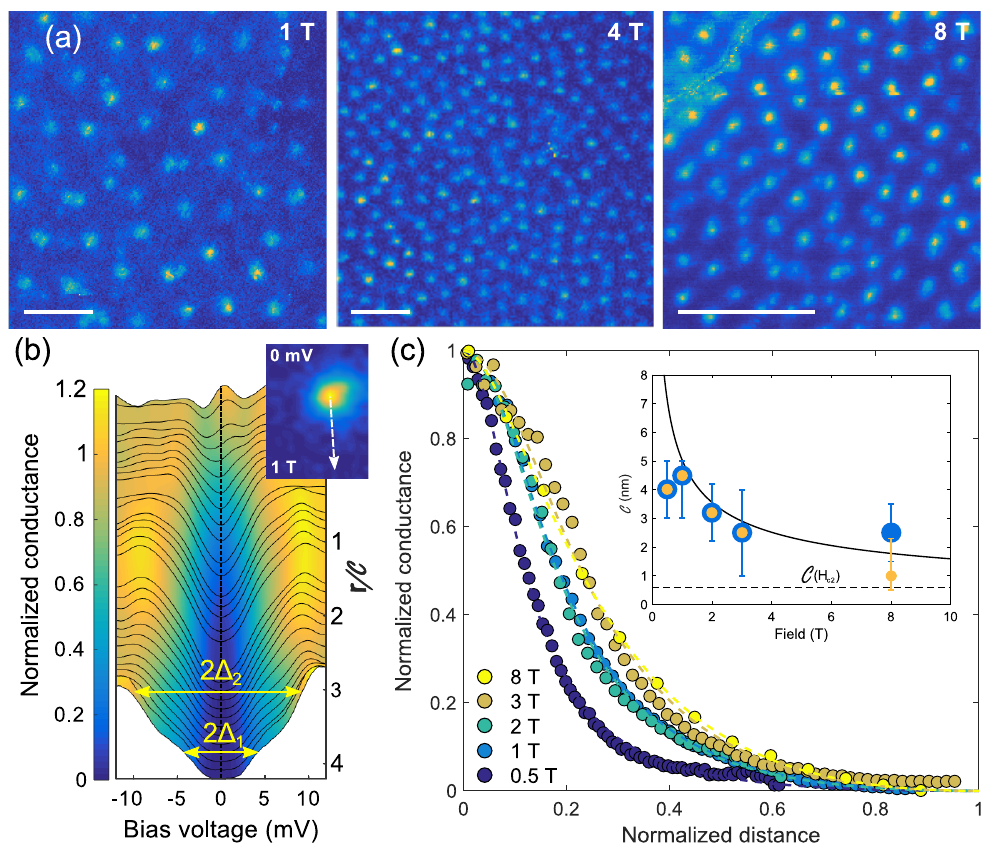}
		\vskip -.1 cm
		\caption{{\bf Vortex lattice and vortex cores of CaKFe$_4$As$_4$.} (a) Zero bias tunneling conductance maps made at different locations and at different magnetic fields. White scale bars are of 82 nm size. The vortex lattice is highly disordered. We show the intervortex distance $a_0$ vs magnetic field in the Appendix, it follows expectations for a hexagonal Abrikosov lattice. (b) The tunneling conductance vs bias voltage from the center of the vortex (top curves) to outside the vortex (bottom curves). We mark the two values of the superconducting gap by yellow arrows. In the inset we show the zero bias conductance map at one vortex, with the path followed in the main panel marked by a white arrow. (c) The normalized conductance vs the distance normalized to the size of the Wigner Seitz unit cell of the vortex lattice for different magnetic fields. The inset shows the vortex core size (obtained as described in the text) vs the magnetic field. Dashed line show the vortex core size extrapolated up to H$_{c2}$.}
		\label{f2}
	\end{center}
\end{figure*}

\begin{figure*}[htb]
	\begin{center}
		\includegraphics[width=0.9\textwidth] {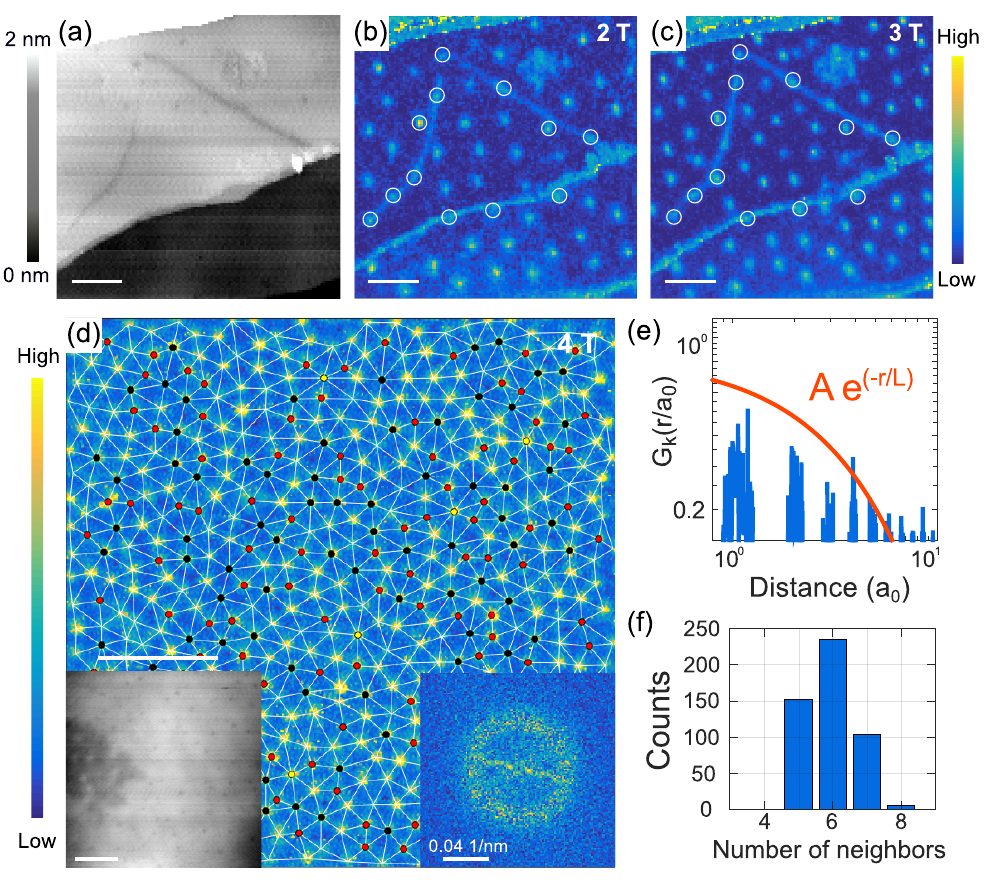}
		\vskip -.1 cm
		\caption{{\bf Disorder in the vortex lattice in CaKFe$_4$As$_4$ due to pair breaking by defects.} (a) STM topographic image with different step edges. There is one large step from bottom left to middle right of the image and two steps at an angle to it. Height differences are given by the grey bar on the left. (b) and (c) The zero bias tunneling conductance map at this location at 2 T (b) and 3 T (c). White scale bars in (a-c) are of 50 nm. White circles mark the position of vortices along the steps in at 2 T. Normalized tunneling conductance scale is given by the right bar. In (d) we show a large zero bias conductance map showing a disordered vortex lattice at 4 T. Vortex positions are Delaunay triangulated (lines and colored dots). Vortices with seven nearest neighbors are marked in black, with five nearest neighbors in red and in yellow for eight nearest neighbors. Bottom left panel shows the corresponding topographic STM image and bottom right panel the Fourier transform of the tunneling conductance image. (e) Shows the positional correlation function vs distance in units of the intervortex distance $a_0=\sqrt{2\phi_0/\sqrt{3}B}$ (top panel) and the histogram over the number of nearest neighbors (bottom panel).}
		\label{f3}
	\end{center}
\end{figure*}

\section{Vortex cores.}

When we apply a magnetic field we observe the vortex lattice over the whole cleaved surface (Fig.\,\ref{f2}(a)). Let us first focus on the shape and size of the vortex core. We show an image with an isolated vortex in the upper panel of Fig.\,\ref{f2}(b). As can be seen there (and also in Fig.\,\ref{f2}(a)), vortex cores are essentially round in CaKFe$_4$As$_4$, showing no conspicuous in-plane anisotropy. The spatial anisotropy of the vortex core is related to the in-plane anisotropy of the superconducting gap \cite{OurReview,Review_Fischer}. We find in-plane isotropic vortex cores, i.e. there is practically no anisotropy in the in-plane gap structure, in agreement with ARPES and our results from quasiparticle interference at zero field.

We show the evolution of the tunneling conductance from the vortex core to the gapped region in between vortices in the main panel of Fig.\,\ref{f2}(b). Results are similar for all magnetic fields up to 8 T. The normalized tunneling conductance at the vortex center shows a small peak for positive bias voltages (upper curves of Fig.\,\ref{f2}(b)). The peak is due to quantized vortex core levels predicted by Caroli, de Gennes and Matricon for clean superconductors \cite{DeGennes-Caroli,OurReview,Review_Fischer}. In many superconductors, a zero bias peak is observed in the tunneling conductance \cite{Hess90,Guillamon08,NbS2}. The peak observed in CaKFe$_4$As$_4$ is slightly asymmetric and located at a non-zero bias. This indicates we are in the so-called extreme quantum limit, where the vortex core states are split with respect to zero bias. As we show with our zero field quasiparticle interference (Fig.\ref{f1}c), the hole band which has its top very close to the Fermi level carries the largest gap $\Delta_2$. This implies that the top of the band, or the Fermi energy of the band, is comparable to the superconducting gap. This shifts the vortex core levels away from zero bias in the extreme quantum limit\cite{Kasahara18112014,doi:10.1143/JPSJ.81.063701,PhysRevLett.77.4074}. The core states are asymmetric with respect to the bias voltage, showing higher intensity for positive bias (empty states).

In Fig.\,\ref{f2}(c) we show the size of the vortex core as a function of the magnetic field, obtained as described in \cite{Fente16}. The vortex core size is treated separately for each group of bands giving different values of the superconducting gap and is defined as ${\cal C}_i \propto \Delta_{OP_{Max},i}(d \Delta_{OP,i} /dr |_{r \to 0})^{-1}$, where $\Delta_{OP,i}$ is the order parameter in each band, $\Delta_{OP_{Max},i}$ the maximum gap between vortices and $r$ the distance from the vortex center. Our fits to ${\cal C}_i$ provide the same values of ${\cal C}_i$ for each band, except at $8\,$T, where we obtain a difference. The extracted ${\cal C}_i$ follow the behavior expected for a superconductor in the clean limit, ${\cal C}_i \propto  1/\sqrt{H}$. When we extrapolate ${\cal C}_i$ to $H_{c2}\approx 70\,$T we find a value for the in-plane coherence length $\xi$ of $0.7\,$nm, which is somewhat smaller, although of order of the value obtained previously from $H_{c2}$ ($1.4\,$nm, using $\xi \approx\sqrt{\phi_0/2\pi H_{c2}}$) \cite{Paul1}. The difference between gap magnitudes is large in CaKFe$_4$As$_4$. But the length scales for the spatial variation of the different components of the order parameter in this two-gap superconductor are the same \cite{PhysRevB.95.064512}.

\section{Vortex lattice and vortex pinning.}

When we measure the vortex lattice, we find six-fold symmetric vortex patterns, indicating that the underlying order is hexagonal over the whole magnetic field range we study. However, the position of vortices is strongly influenced by the order parameter suppression observed around some defects. As we have shown in Fig.\,\ref{f1}(c) at zero field, there are some locations where we observe pair breaking in form of a finite zero bias conductance. In some fields of view, these locations are not related to features in the topography (as in Fig.\,\ref{f1}(c)), but in other fields of view they can be linked to a step on the surface. In Fig.\,\ref{f3}(b)-(c) we show tunneling conductance maps taken at $2\,$T and at $3\,$T at the same location. Images show several steps producing lines with a finite zero bias conductance. Remarkably, some vortices are positioned at these lines and some even form a chain along the step with a high zero bias conductance. The step might be due to a structural defect in the bulk that suppresses the order parameter. Vortices around the defect change their position when increasing the magnetic field from $2\,$T to $3\,$T, but the vortices marked by white circles remain at the same positions, along the lines with a high zero bias conductance. Thus, vortices are pinned to the locations with finite zero bias conductance. As can be seen in Fig.\,\ref{f3}(b)-(c), the lateral size of the locations with pair breaking coincides with the vortex core size ${\cal C}_i$.

In Fig.\,\ref{f3}(d) we discuss a map showing many vortices. We have Delaunay triangulated vortex positions. The concentration of defects in the vortex lattice is very high, with roughly the same number of five-fold and seven-fold vortices, that sums up to nearly the number of vortices with six nearest neighbors (Fig.\,\ref{f3}(f)). The positional correlation function $G_K(r)$ (Fig.\,\ref{f3}(e)) decays following $e^{-r/L}$ with $L$ of 4-5 intervortex distances $a_0$. We see that ordered bundles include just a few hexagons. The behavior in Fig.\,\ref{f3}(a)-(c) is characteristic for isolated vortex pinning, i.e., vortices are located exactly on the pinning centers. The behavior in Fig.\,\ref{f3}(d) is characteristic for collective pinning, i.e. vortices distribute on a disordered hexagonal lattice, being probably pinned to locations with order parameter suppression lying well below the surface \cite{Blatter94}.

Disorder in the vortex lattice has been studied by STM in several iron based superconductors. Experiments on doped compounds report varied behavior. Vortex lattice images on (Sr$_{0.75}$K$_{0.25}$)Fe$_2$As$_2$ \cite{Song13} and Ba(Fe$_{0.9}$Co$_{0.1}$)$_2$As$_2$ \cite{Hoffman2} show a disordered vortex lattice while (Ba$_{0.6}$K$_{0.4}$)Fe$_2$As$_2$ \cite{Shan11} presents a more ordered lattice for the same values of the magnetic fields. This indicates that charge doping by itself does not explain the origin of vortex pinning in these materials. Instead, it has been suggested that electronic inhomogeneity induced by ion size mismatch between substituted elements accounts for the different pinning strengths found in these compounds (for instance, the mismatch between K and Ba is five times smaller than between K and Sr) \cite{Song13}. In bulk Fe(Se,Te) irradiated by heavy ions, vortices are pinned to the amorphous regions at the columnar defects created by irradiation, leading to a complicated pinning landscape \cite{Massee15}. In FeSe, vortices are preferentially positioned near twin boundaries where superconductivity is strongly suppressed \cite{Song12}. In the stoichiometric pnictide LiFeAs a very disordered vortex lattice has been observed in absence of twin boundaries or substitutional disorder but at much larger values of $H/H_{c2}$, for which vortex disorder due to lattice softening is much more significant than here \cite{Hanaguri12}.

The present results in stoichiometric CaKFe$_4$As$_4$ uniquely show the link between the locations with a finite density of states at the Fermi level and the vortex positions. Studying a stoichiometric material has been essential to obtain this conclusion. Our experiment shows that the disordered vortex lattice characteristic of FeBSC superconductors is due to order parameter suppression at defects whose size is of the order of the vortex core size.

\section{Discussion and conclusions.}

From the structural point of view, the CaKFe$_4$As$_4$ system is related to CaFe$_2$As$_2$ but has alternating Ca and K layers\cite{Hirschfeld11}. In the latter case, the Fe-As sheets are centered between the Ca-planes. In CaKFe$_4$As$_4$ the larger K ions shift these sheets towards K and destroy the n-glide symmetry across the Fe plane present in most iron based superconductors. Moreover, the As atoms next to K and Ca become inequivalent and adopt different distances from the Fe plane. Band structure calculations in CaKFe$_4$As$_4$ show that the band dispersion has a much stronger two-dimensional character than in the CaFe$_2$As$_2$ and related systems, which can enhance Coulomb interactions and favor unconventional $s\pm$ superconductivity\cite{PhysRevB.96.094521}. Calculations indeed show that order parameter changing sign between electron and hole pockets is favored by antiferromagnetic spin fluctuations and Coulomb repulsion\cite{PhysRevB.96.094521}. The structure of sign changes is expected to be more intricate with less Coulomb repulsion, showing sign changes of the order parameter within hole and electron bands. Here we show that there are no sign changes within the hole bands, or otherwise the quasiparticle interference scattering we consider (Fig.\,\ref{f1}), would lead to suppression of superconductivity.

Along the same line, recent neutron scattering experiments measure the dynamical spin susceptibility and find a feature at the nesting vector between electron and hole bands, located at an energy of 12.5 meV\,\cite{doi:10.7566/JPSJ.86.093703}. In our data we can identify a small shoulder in the quasiparticle interference at this energy (see Fig.\,\ref{f1}(b)), although it seems quite adventurous to claim a strong relationship with neutron scattering experiments. In any event, the resonance in the dynamical spin susceptibility is actually quite broad and develops at low temperatures within the superconducting phase, so that we do not expect a clear signal in the quasiparticle interference of the hole bands. However, taken altogether, neutron scattering highlights the relevance of spin fluctuations and our data shows that the sign change is between electron and hole bands. This establishes CaKFe$_4$As$_4$ as a paradigmatic $s\pm$ spin-fluctuation mediated superconductor.

Another interesting feature of CaKFe$_4$As$_4$ is that the hole bands reach their top at around 30-40 meV from the Fermi level. This provides a ratio of $\Delta/E_F\approx 0.25$ and shifts the lowest lying Caroli de Gennes Matricon vortex core level (lying at $\Delta^2/2E_F$\cite{Hayashi98}) from $E_F$ (Fig.\,\ref{f2}(b)). The value of $\Delta/E_F$ found here in CaKFe$_4$As$_4$ is below the values which place FeSe and related systems in proximity of the crossover between Bardeen-Cooper-Schrieffer to Bose-Einstein-Condensation (BCS-BEC)  \cite{Kasahara18112014,Okazaki2014}. However, it's of the same order ($\Delta/E_F\approx 1$ for FeSe or 0.6 for Fe(Se,Te)\cite{Kasahara18112014,Okazaki2014}).

In turn, this also shows that small changes by doping can significantly influence the superconducting and normal state properties. It would be interesting to search for doping induced modifications in the superconducting order parameter sign between the electron and hole bands. If the hole band can be brought closer to the Fermi level, predictions range from intraband $s\pm$ superconductivity to time-reversal symmetry breaking $s+is$ superconductivity, with superconducting order parameter phase differences among bands that are not multiples of $\pi$, due to the competition between Coulomb repulsion and spin fluctuations\cite{PhysRevB.96.014517,HIRSCHFELD2016197,PhysRevLett.114.107002}. Furthermore, the presence of hedgehog spin vortex crystal in Ni doped CaKFe$_4$As$_4$, which could be related to the absence of glide plane \,\cite{PhysRevB.96.014517,PhysRevB.95.075104,PhysRevB.93.014511,Meier17}, adds an interesting ingredient likely favoring unconventional superconducting properties. Stoichiometric CaKFe$_4$As$_4$ is an advantageous starting point for exploring these ideas compared with other iron based superconducting systems.

Finally, we note that there is no evidence for a significant in-plane anisotropy in CaKFe$_4$As$_4$. The situation is quite different to the four-fold anisotropy observed in vortex cores in LiFeAs or the strong two-fold anisotropy of vortices in FeSe \cite{Hanaguri12,PhysRevB.85.020506,Song2011}, but it is similar to the situation found in Fe-As based systems such as for instance Ba$_{0.6}$K$_{0.4}$Fe$_2$Se$_2$\cite{Shan11,Hoffman2,HoffmanPnictides}. Maybe the structural stability of the CaKFe$_4$As$_4$, which is much less pressure or strain sensitive than related compounds, suffers from less in-plane anisotropy. In turn, this implies that the in-plane anisotropy in the bandstructure and the structural transitions are key ingredients influencing the behavior of other iron-based superconductors.

\section{Acknowledgments}

We acknowledge discussions with R. Prozorov and S. Vieira. Work done in Madrid (AF, IG and HS) was supported by the Spanish Ministry of Economy and Competitiveness (FIS2014-54498-R, MDM-2014-0377), by the Comunidad de Madrid through program Nanofrontmag-CM (S2013/MIT-2850), by the European Research Council PNICTEYES grant agreement 679080 (IG) and by FP7-PEOPLE-2013-CIG 618321, Cost CA16218, Nanocohybri) and Axa Research Fund. SEGAINVEX-UAM and Banco Santander are also acknowledged. Work done in Ames Lab (PCC, SLB, VGK, WRM and TK) was supported by the U.S. Department of Energy, Office of Basic Energy Science, Division of Materials Sciences and Engineering. Ames Laboratory is operated for the U.S. Department of Energy by Iowa State University under Contract No. DE-AC02-07CH11358. WRM was supported by the Gordon and Betty Moore Foundation's EPiQS Initiative through Grant No. GBMF4411.

\newpage

\section{Appendix}

\subsection{Crystal structure and absence of the surface reconstruction observed in CaFe$_2$As$_2$ and related systems}

As discussed above, the crystal structure of CaKFe$_4$As$_4$, holds relevant differences to CaFe$_2$As$_2$, such as the absence of a glide plane and the positions of the Fe-As layers. However, they are both tetragonal phases with quite similar atomic positions in them \cite{Paul1}. It is, therefore, reasonable to assume that the cleaving in CaKFe$_4$As$_4$ occurs next to Fe-As blocks as happens in CaFe$_2$As$_2$ and its doped compounds. These often show a $2\times 1$ surface reconstruction consisting of rows of Ca atoms that remain on the surface after cleaving \cite{HoffmanPnictides,Massee09,Shan11,Fente16b}. We do not observe such a reconstruction in the surfaces of CaKFe$_4$As$_4$. The notable stress sensitivity of CaFe$_2$As$_2$ is attributed to the small size of the Ca ions. The different size and regular order of Ca and K may mitigate these steric effects in CaKFe$_4$As$_4$\,\cite{Ran11}. We believe that the released strain in CaKFe$_4$As$_4$ influences the structure of the top-most surface in CaKFe$_4$As$_4$ producing atomically flat surfaces and no surface reconstruction.

\subsection{Quasiparticle interference and symmetrization}

\begin{figure}[htb]
	\begin{center}
		\includegraphics[width=0.45\textwidth] {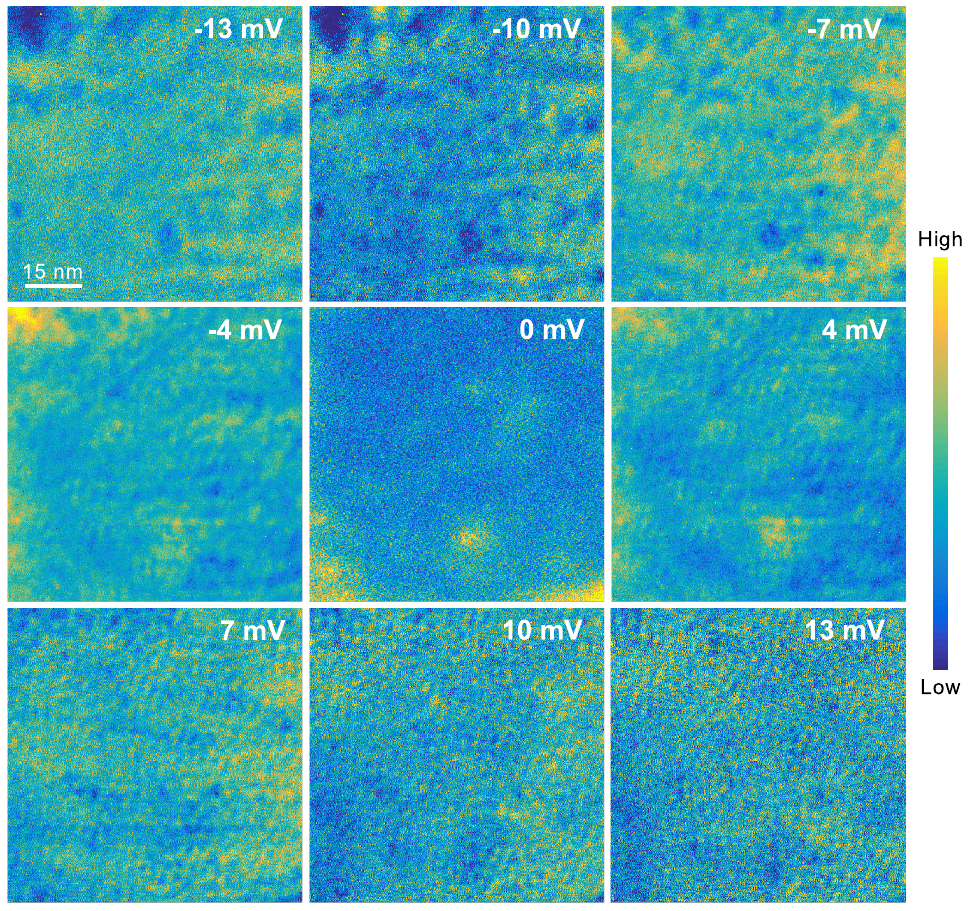}
		\vskip -.1 cm
		\caption{Conductance maps as a function of energy showing the evolution of the QPI pattern analyzed in Fig.\,\ref{f1}.}
		\label{f5}
	\end{center}
\end{figure}

We present in Fig.\,\ref{f5} the real space conductance maps of our quasiparticle interference experiments. Background modulation evolves with energy producing the reciprocal space images seen in the left panel of Fig.\,\ref{f1}. No modulation is seen for energy values inside the superconducting gap.

\begin{figure}[htb]
	\begin{center}
		\includegraphics[width=0.45\textwidth] {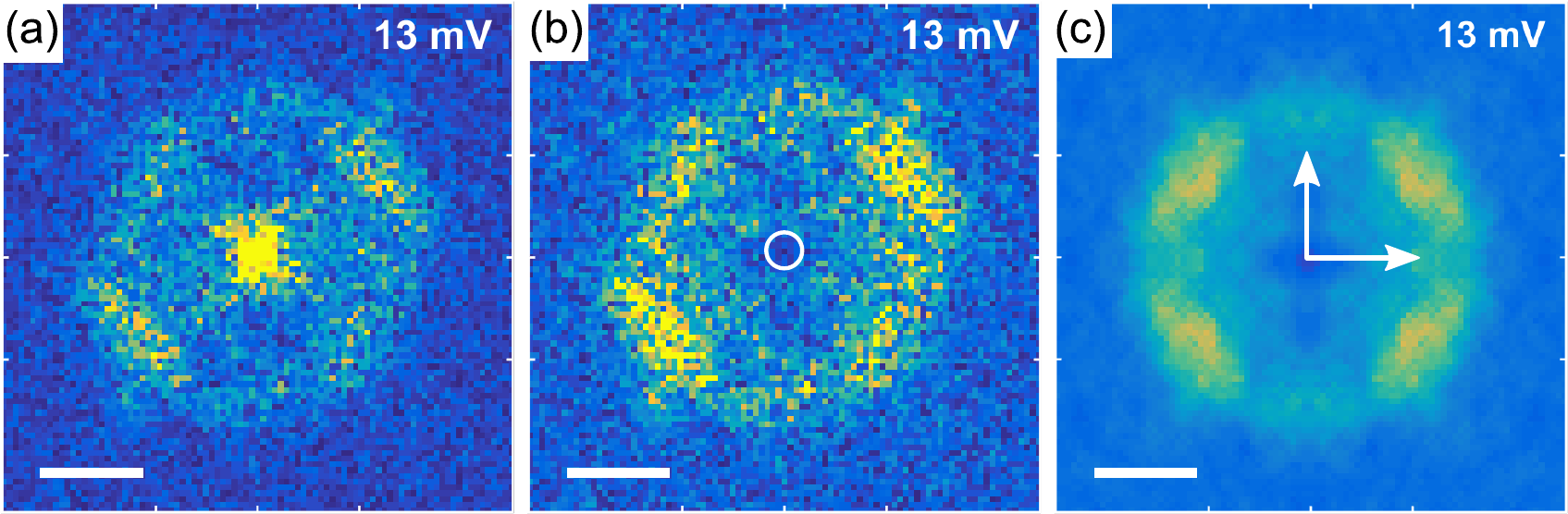}
		\vskip -0.1 cm
		\caption{In (a) we show the raw Fourier transform of our conductance map at 13 mV. In (b) we have removed the central circle marked in white symbol. In (c) we show the result of a Gaussian smooth to (b). White scale bar in FFT is $0.1$ $\pi/a$ long.}
		\label{fSymQPI}
	\end{center}
\end{figure}

In Fig.\,\ref{fSymQPI} we show the process followed to obtain the Fourier transform of the tunneling conductance maps. We discuss the Fourier transform of the map at 13 mV as a representative example. In Fig.\,\ref{fSymQPI}(a) we show the Fourier transform of raw data. In Fig.\,\ref{fSymQPI}(b) we remove the central peak and in Fig.\,\ref{fSymQPI}(c) we show the result of the symmetrization. The raw data show some asymmetry due to spatial structure of defects. The symmetry of the in-plane lattice is expected to be $4mm$. However, we decided to apply a $2mm$ symmetrization. This leaves part of the asymmetry due to defect scattering, but the figures are more clear and it provides a representation closer to the raw data. In any event, this does not influence the results, which focus on the bias voltage dependence of the radial average of the intensity of the quasiparticle pattern along the path marked by a white dashed line, which corresponds to the circle of maximum intensity in Fig.\,\ref{f1}.

\begin{figure}[htb]
	\begin{center}
		\includegraphics[width=0.45\textwidth] {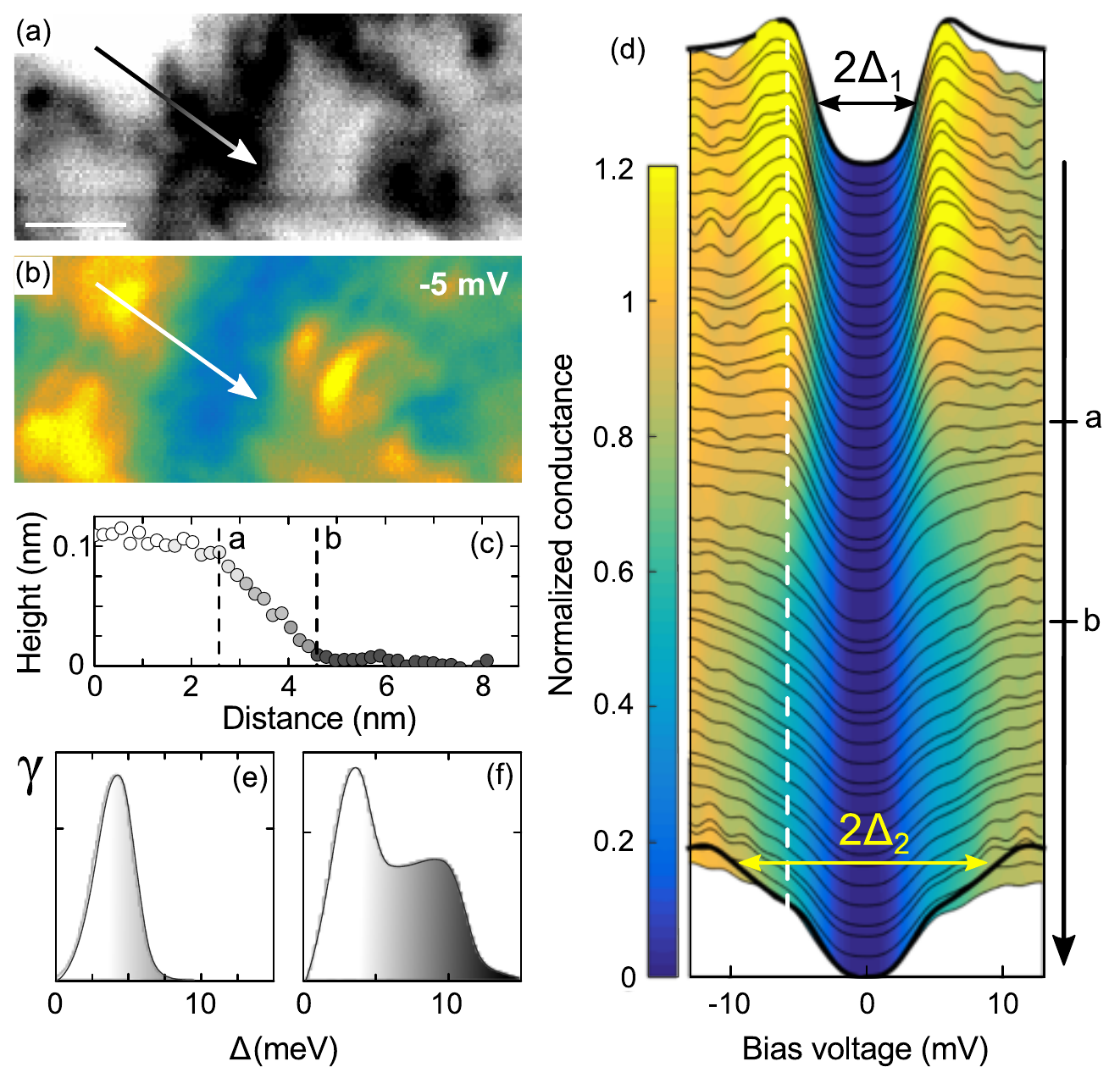}
		\vskip -0.1 cm
		\caption{(a) STM topography of an almost atomically flat area. The difference between black and white is given by the height profile in (c) along the arrow. The white scale bar is $4\,$nm long. (b) Represents the normalized tunneling conductance over the are in (a) taken at $-5\,$meV. The image shows a clear relation between the gap structure and the part of the surface in which we are tunneling. (c) Height profile along the line scan shown in (a). (d) Normalized conductance curves taken along the arrow in (a) and (b). We mark by arrows the main values of the superconducting gap found, as discussed in the text, $\Delta_1$ and $\Delta_2$. (e,f) Distribution of gap values $\Delta$ being the gap values and the relative weight of the corresponding gap values  $\gamma$. Using the distributions shown in the figure, we generate a superconducting density of states summing over all gaps with the corresponding weights (see text). We convolute the resulting density of states with the derivative of the Fermi function to find the tunneling conductance given by the black lines at the top and bottom of the panel in (d).}
		\label{f8}
	\end{center}
\end{figure}

\subsection{Tunneling conductance vs position in different tunneling planes}

In Fig.\,\ref{f8} we show tunneling spectroscopy in a very small and flat area. We obtain tunneling conductance curves with two characteristic gap values, $\Delta_1 = 3$\,meV $= 0.6\,\Delta_0$ and $\Delta_2 = 8$\,meV $= 1.5\,\Delta_0$ (with $\Delta_0 = 1.76\,k_B T_c$). To determine these values, we calculate the tunneling conductance assuming a density of states of the form $\sum_{\Delta_i}{\gamma_i\,Re\left(\sqrt{\frac{E}{E^2-\Delta_i^2}}\right)}$ and convolute the result with the derivative of the Fermi function to include the effect of temperature. $\Delta_1$ and $\Delta_2$ are the peaks thus obtained in the distribution of weights $\gamma_i$. The gap values we obtain agree with previous STM and penetration depth experiments \cite{Prozorov16,Kaminski16}. The weight of $\Delta_1$ and $\Delta_2$ in the tunneling conductance changes when crossing a stripe, leading to different distribution of $\gamma_i$ (Fig.\,\ref{f8}\,(e,f)). The contribution of different parts of the Fermi surface to the tunneling conductance changes as a function of the atomic plane into which we are tunneling, because in each surface the tunneling matrix elements, linked to the terminating surface atoms, are different \cite{Shan11, HoffmanPnictides, OurReview}. Very similar results are obtained in (Ba$_{1-x}$K$_x$)Fe$_2$As$_2$\,\cite{Shan11}.

\subsection{Obtaining the vortex core radius $\cal C$}

To obtain the magnetic field dependence of the vortex core radius $\cal C$, we take zero bias conductance maps of vortices that have a round shape and are far from pinning centers. We then center the image at a single vortex and make the angular average of the tunneling conductance for each distance $r$ from the vortex center. We define the normalized conductance  as
\begin{equation}
\sigma =\frac{\sigma_0(r)-\sigma_0(r^*)}{\sigma_0(0)-\sigma_0(r^*)}\,.
\label{eq1}
\end{equation}
where $\sigma$ is the normalized conductance shown in the Fig.\,\ref{f2}(c), $\sigma_0$ the angular average over the center of the vortex in a tunneling conductance map, $r$ the distance from the vortex center. $r^*$ is the distance from the vortex center to the center of a vortex triangle, or the size of the Wigner Seitz cell of the vortex lattice. This is also the normalization parameter used in Fig.\,\ref{f2}(c). Further details are given in Ref.\cite{Fente16}.
\\

\subsection{Vortex lattice at different fields}

\begin{figure}[htb]
	\begin{center}
		\includegraphics[width=0.45\textwidth] {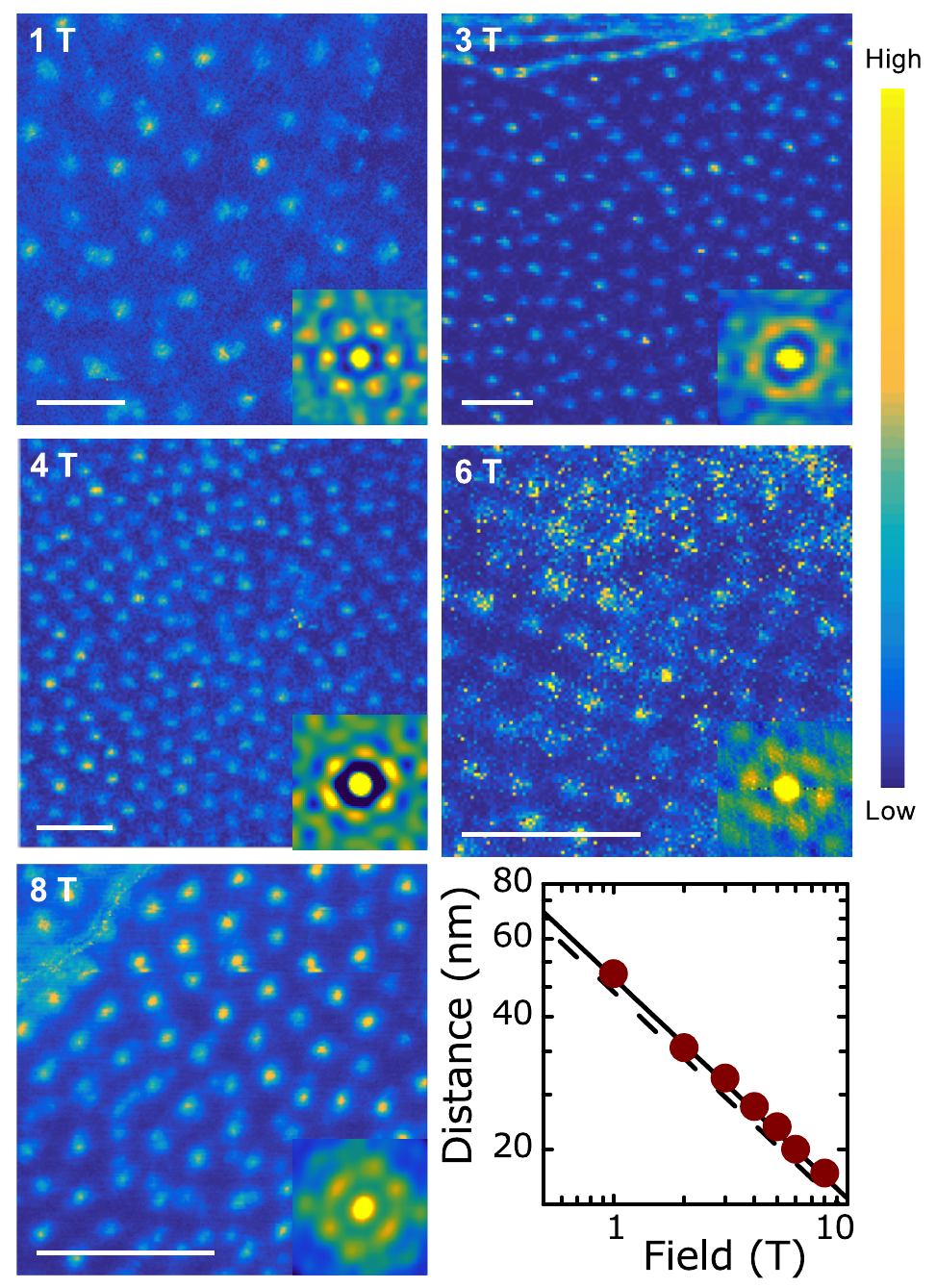}
		\vskip -0.1 cm
		\caption{Zero bias conductance maps in fields of view between $1$\,T and $8$\,T taken at different positions. The color bar on the right provides the zero bias normalized tunneling conductance. The white bar in the images is of $82$\,nm size. The insets in the images show the autocorrelation function of the images. Note that the lattice remains hexagonal, within first nearest neighbors, over the whole magnetic field range. The lower right panel shows the intervortex distance, obtained from the autocorrelation function, vs the magnetic field and the black lines are the expected intervortex distance for hexagonal (solid line) and square (dashed line) lattices, corresponding to $a_0\propto\frac{1}{\sqrt{H}}$.}
		\label{f6}
	\end{center}
\end{figure}

Fig.\,\ref{f6} contains zero bias conductance maps taken in different areas at different fields between $1$\,T and $8$\,T. Their autocorrelation function is included as inset in each map. It shows six-fold features as expected within a triangular lattice with disorder. The magnetic field dependence of the intervortex distance $a_0$ follows expectations for a triangular vortex lattice (solid line), $a_0^2=  2\Phi_0/ \sqrt{3}H$ (where $\Phi_0$ is the flux quantum).

Fig.\,\ref{f6} also shows how vortex position follow pair breaking defects. See for example the $3$\,T image, where the vortices align following a curved line with a finite zero bias conductance in between vortices going from the top left part to the center of the image.


%

\end{document}